%% file: paper_hepex.tex
\documentclass[twocolumn,showpacs,aps,prl,superscriptaddress]{revtex4}

\usepackage{graphicx}
\usepackage{dcolumn}
\usepackage{amsmath}
\usepackage{epsfig}

\input{bsg_symbols}

\newcommand{\BABARPubYear}    {06}
\newcommand{\BABARPubNumber}  {037}

\newcommand{\SLACPubNumber} {12002}

\def\figurebox#1#2#3{%
    \def\arg{#3}%
    \ifx\arg\empty
    {\hfill\vbox{\hsize#2\hrule\hbox to #2{\vrule\hfill\vbox to #1{\hsize#2\vfill}\vrule}\hrule}\hfill}%
    \else
    {\hfill\epsfbox{#3}\hfill}%
    \fi}

\begin{document}

\preprint{\babar-PUB-\BABARPubYear/\BABARPubNumber} 
\preprint{SLAC-PUB-\SLACPubNumber}

\begin{flushleft}
\babar-PUB-\BABARPubYear/\BABARPubNumber\\
SLAC-PUB-\SLACPubNumber\\
\end{flushleft}

\title{
{\large Measurement of the Branching Fraction and Photon Energy Moments of \bxsg and
$\acp(B \to X_{s+d}\gamma)$} 
}

\input authors

\date{\today}

\begin{abstract}
The photon spectrum in \bxsg decay, where \xs is any strange hadronic state, is studied  using a data sample 
of $88.5\times10^{6}$ $\epem \to \Y4S \rightarrow \BB$ decays collected by the \babar\ experiment at SLAC. 
The partial branching fraction, $\Delta\BR(\bxsg)=(3.67\pm0.29(stat.)\pm0.34(sys.) \pm0.29(model)) \times 10^{-4}$, 
the first moment $\efmbrest=2.288\pm0.025\pm0.017\pm0.015  \gev$ and the second moment $ \esmbrest 
=0.0328\pm0.0040\pm0.0023\pm0.0036 \gev^{2}$ are measured for the photon energy range  $1.9 \gev < \eg < 2.7 \gev$. 
They are also measured for narrower \eg ranges. The moments are then fit to recent theoretical calculations 
to extract the Heavy Quark Expansion  parameters, \mb and \mupisq, and to extrapolate the partial branching 
fraction to $\eg>1.6 \gev$. In addition, the direct $\CP$ asymmetry $\acp(B \to X_{s+d}\gamma)$ is measured to 
be $-0.110\pm0.115(stat.)\pm0.017(sys.)$.

\end{abstract}

\maketitle

In the Standard Model (SM) the  radiative decay of the \b quark,  \bsg , proceeds via a loop diagram, and is 
sensitive to possible new physics, with new heavy particles participating in the loop \cite{ref:new-physics}. 
Next-to-leading-order  SM calculations for the branching fraction give $\BR(\bxsg) = (3.61^{+0.37}_{-0.49}) 
\times 10^{-4}$\,($\eg > 1.6\gev$) \cite{ref:br-theory}, and calculations to higher order, which are expected 
to considerably decrease the uncertainty,  are currently underway \cite{ref:theory-future}. The shape of the 
photon energy spectrum, which is insensitive to non-SM physics~\cite{ref:KN},  can be used to determine the
Heavy Quark Expansion (HQE)  parameters, \mb and \mupisq ~\cite{ref:BBU,ref:N},  related to the mass and momentum 
of the  \b quark within the $B$ meson. These parameters can be used to reduce the error in the extraction of the 
CKM matrix elements $\Vcb$ and $\Vub$ from semi-leptonic  \B-meson decays~\cite{ref:henning}.  New physics can 
also significantly enhance the direct \CP asymmetry for \bsg and \bdg decay\cite{ref:br-theory},
$
  \acp = \frac{\Gamma(\b\to\s\g + \b\to\d\g)-
       \Gamma(\bbar\to\sbar\g + \bbar\to\dbar\g)}
       {\Gamma(\b\to\s\g + \b\to\d\g)+
       \Gamma(\bbar\to\sbar\g + \bbar\to\dbar\g)}
$
which is $\approx 10^{-9}$ in the SM \cite{ref:acppredict}. Measurements of this joint asymmetry complement those of
\acp in \bsg \cite{ref:acpexperiment} to constrain new physics models.

This letter reports on a fully-inclusive analysis of $\bxsg$ decays  collected from $\epem\to\Y4S\to\BB$, where 
the photon from the decay of one \B meson is measured, but the $X_s$ is not reconstructed.  This avoids incurring 
large uncertainties from the modeling of the $X_{s}$ fragmentation, but at the cost of high backgrounds which need 
to be strongly suppressed. The principal backgrounds are from  other \BB decays containing a  high energy photon 
and from continuum \qqbar\ ($q=udsc$) and $\tau^{+}\tau^{-}$ events. The continuum background, including a 
contribution from initial state radiation (ISR), is suppressed principally by requiring a high-momentum lepton from
the non-signal \B decay,  and also by discriminating against its more jet-like topology. The \BB  background to high 
energy photons, dominated by $\piz$ and $\eta$ decays, is reduced by vetoing on reconstructed $\piz$ or $\eta$.
The residual continuum background is subtracted using off-resonance data taken at a center-of-mass  energy 40\mev  
below that of the \Y4S, while the  remaining \BB background is estimated  using a Monte Carlo simulation which has 
been checked and corrected using data control samples. Previous inclusive measurements of $\bxsg$ have been
presented by the CLEO~\cite{ref:cleobsg}, BELLE~\cite{ref:bellebsg} and \babar\ ~\cite{ref:semi-inclusive} 
collaborations using alternative techniques which incur different systematic uncertainties.

The results presented are based on data collected with the \babar\ detector~\cite{ref:babar-nim} at the PEP-II
asymmetric-energy \epem collider located at the Stanford Linear Accelerator Center. The on-resonance integrated
 luminosity is 81.5\invfb, corresponding to 88.5 million \BB events. Additionally, 9.6\invfb of off-resonance data 
are used in the continuum background subtraction. The \babar\ Monte Carlo simulation program, based on 
GEANT4~\cite{ref:geant}, EVTGEN~\cite{ref:evtgen} and JETSET~\cite{ref:jetset}, is used to generate samples of 
\BpBm and \BzBzb (excluding signal channels), \qqbar, $\tau^{+}\tau^{-}$, and signal events. The signal models 
used to calculate efficiencies are based on references~\cite{ref:BBU}~(``kinetic scheme'') and 
~\cite{ref:N}~(``shape function scheme'') and on an earlier calculation~\cite{ref:KN}(``KN''). These predictions 
approximate the \xs resonance structure with a smooth distribution in  \mxs. This is reasonable except at the lowest 
masses where the $\Kstar(892)$ dominates the spectrum. Hence the portion of the \mxs spectrum below 1.1\gevcc is 
replaced by a Breit-Wigner $\Kstar(892)$ distribution. The analysis was done ``blind'' in the range of reconstructed 
photon energy \egcms from 1.9 to 2.9\gev (asterisk denotes the \Y4S rest frame); that is, the on-resonance data were 
not looked at until all selection requirements were set and  the corrected backgrounds determined. The signal range 
is limited by high \BB backgrounds at low \egcms .

The event selection begins by  finding at least one photon candidate  with, $1.6<\egcms<3.4\gev$, in the event. 
A photon candidate is a localized electromagnetic calorimeter energy deposit with a lateral profile consistent 
with that of a single photon. It is required to be isolated by 25 \cm from any other energy deposit and to be 
well contained in the calorimeter ($-0.74 < \cos \theta_\gamma < 0.93$), where $\theta_\gamma$ is the polar angle 
with respect to the beam-axis. Photons that are consistent with originating from an identifiable \piz or 
$\eta\to\gamma\gamma$ decay are vetoed.  Hadronic events are selected by requiring at least three  reconstructed 
charged particles and the normalized second Fox-Wolfram moment  $R_2^*$ to be less than 0.55. To reduce radiative 
Bhabha and two-photon  backgrounds, the number of charged particles plus half the number of photons with energy 
above 0.08\gev is required to be $\ge 4.5$.

Event shape variables are used to exploit the difference in topology of isotropic \BB events and jet-like continuum
 events. This is accomplished by the  $R_2^*$ requirement as well as a single linear discriminant  formed from 
nineteen different variables.  Eighteen of the quantities are the sum of charged and neutral energy found in 
10-degree cones (from 0 to 180 degrees) centered on the photon candidate direction; the photon energy is not 
included. Additionally the discriminant includes  $R_2^\prime/R_2^*$, where $R_2^\prime$ is the normalized second 
Fox-Wolfram moment calculated in the frame recoiling against the photon, which for ISR events is the \qqbar\ rest 
frame. The discriminant coefficients were determined by maximizing the separation power between simulated signal
and continuum events.

Lepton tagging further reduces the backgrounds from continuum events. About 20\% of \B mesons decay semi-leptonically 
to either $e$ or $\mu$. Leptons from hadron decays in continuum events tend to be at lower momentum. Since the tag 
lepton comes from the recoiling \B meson, it does not compromise the inclusiveness of the \bxsg\ selection. The tag 
lepton is required to have momentum $p^{*}_{e}>1.25\gevc$ for electrons and $p^{*}_{\mu}>1.5\gevc$ for muons. 
Additionally requiring the photon-lepton angle, $\cos\theta^{*}_{\gamma\ell} > -0.7$  removes more continuum 
background, in which the lepton and photon candidates tend to be back-to-back. Finally the presence of a relatively
high-energy neutrino in semi-leptonic \B decays is exploited by requiring the missing energy of the event, 
$E^{*}_\mathrm{miss}>0.8\gevc$. Virtually all of the tagging leptons arise from the decay $B\to
X_c\ell\nu$. The rate of such events in the  simulation is corrected as a function of lepton 
momentum~\cite{ref:BBsemilep}.

The event selection is chosen to maximize the statistical significance of the expected signal using simulated 
signal (KN with \mb=4.80 \gevcc , \mupisq = 0.30 $\gev^{2}$) and background events, allowing for the low statistics 
of the off-resonance data used for the subtraction of continuum background. After selection the low energy 
range, $1.6 < \egcms < 1.9 \gev$, is dominated by the \BB background, while the high energy range, 
$2.9 < \egcms < 3.4\gev$, is dominated by the continuum background; they provide control regions for the \BB 
subtraction and continuum subtraction, respectively. The signal region lies  between 1.9\gev and 2.7\gev. The 
signal efficiency ($\approx 1.6\%$ for this \egcms range) depends on \egcms and the signal model, but has negligible 
dependence on the details of the fragmentation of the $X_{s}$. 

\input results.tex

The \BB background is estimated with the simulated \BB data set. It consists predominantly of photons originating 
from $\piz$ or $\eta$  decays ($\approx 80\%$). Other significant sources are $\antineutron$'s which fake photons by 
annihilating in the calorimeter and electrons that are misreconstructed or lost,  or that undergo hard Bremsstrahlung.
 The \pizeta\ background simulation is compared to data by using the same selection criteria as for $\bxsg$ but 
removing the \pizeta\ vetos. The photon energy and lepton momentum thresholds are relaxed  to $\egcms > 1.0 \gev$, 
$ p^{*}_{e}>1.0 \gevc $, $ p^{*}_{\mu}>1.1 \gevc$ to gain statistics.  The yields of \pizeta\ are measured in bins 
of $E^{*}_{\pizeta}$ by fitting the $\gamma\gamma$ mass distributions in on-resonance data, off-resonance data and
simulated \BB background. Correction factors to the \piz ($\eta$) components of the \BB simulation are derived from 
these yields, including a small adjustment for the different efficiencies of the \piz ($\eta$) vetoes between data 
and simulation. As no $\antineutron$ control sample could be isolated, this source of \BB background is corrected 
by comparing in data and simulation the  inclusive $\antiproton$ yields in \B  decay and the calorimeter response 
to $\antiproton$'s, using a $\overline{\Lambda}\to \overline{p}\pip$ sample.  The electron component of the \BB  
simulation is corrected  with electrons from a  Bhabha data  sample, taking into account the lower track multiplicity
 of these events compared to the signal events. Finally, the small contributions from $\omega$ and $\etapr$ decays 
are corrected using  inclusive \B decay data. After including all corrections and systematic errors  the expected 
background yield from the simulation in the \BB control region ($1.6 < \egcms < 1.9 \gev$) is $1667 \pm 54$ events, 
compared to  $1790 \pm 64$ events observed in data after continuum subtraction. Note that a small contribution in 
this region from the expected signal ($\approx 20$ to $40$ events) has been neglected in this comparison. In the high
 energy control region $2.9 <\egcms<3.4 \gev$ the expected background is $390  \pm 20$ events, compared to  
$393\pm58$ events observed in data.

\begin{figure}[htb]
\begin{center}   
  \includegraphics[width=.5\textwidth]{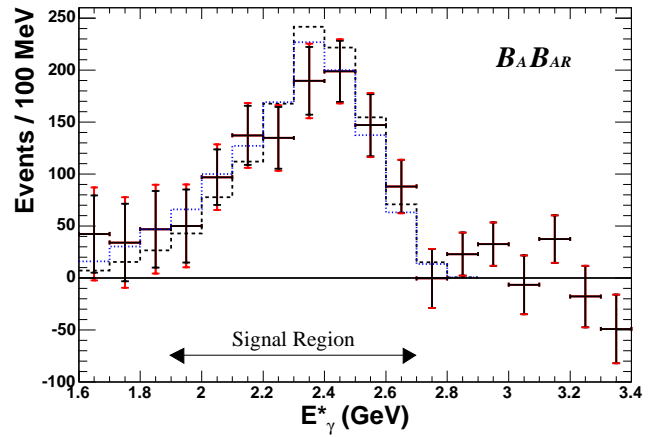}   
\end{center}
\vspace{-0.2in}
\caption{The photon energy spectrum after background subtraction, uncorrected for efficiency. The inner error bars 
are  statistical and the outer include systematic errors added in quadrature. The histograms show the spectra for
 values of \mb and \mupisq from the best fits to the moments in the kinetic scheme (dashed) and shape function 
scheme (dotted), normalized to the data in the signal region.}
\label{fig:spectrumuncorrected} 
\end{figure}

Figure~\ref{fig:spectrumuncorrected}  shows the measured spectrum for signal and control regions  after the  
\BB and continuum backgrounds have been subtracted.  To extract partial branching  fractions (PBFs) and first and
second moments from this spectrum it is necessary to first correct for efficiency. Theoretical predictions are made 
for the true \eg in the \B meson rest frame, whereas the experimental  measurements are made with reconstructed 
\egcms in the \Y4S frame. Hence it is also necessary to  correct for smearing due to the asymmetric calorimeter 
resolution and the Doppler shift between the \Y4S frame and the \B rest frame. The  efficiency and smearing 
corrections depend upon the assumed signal model (underlying theory and parameter values). In a broad selection  
of signal models it is found that the efficiency for each \egcms range has a model-independent linear relationship 
to the mean \egcms in that range. Hence a nominal signal model is chosen for which the mean matches the data, and 
a model-dependence uncertainty is assigned to the PBFs and moments based on signal models within one (statistical 
and systematic) standard deviation of the measured mean \egcms.  To correct for resolution smearing a small 
multiplicative correction to the PBF and small additive corrections to the first and second moments are computed 
using the nominal signal model, and an uncertainty assigned based on a conservative range of models. The 
model-dependence uncertainty from the smearing correction is fully correlated with the corresponding uncertainty 
of the efficiency correction. 

The results for four energy ranges are given in Table~1 along with the statistical, systematic and model errors. 
The PBFs have been corrected to exclude a $(4.0\pm0.4)\%$~\cite{ref:CKMfitter,ref:br-theory} contribution from $\bdg$.
The systematic errors are described below and  the associated correlation matrices are given in 
the appendix.

The most significant systematic uncertainty in the measurement of the spectrum is from the uncertainty in the 
corrections to the $\BB$ background simulation.  It is due mostly to the statistical uncertainty on the correction 
factors derived from the $\pizeta$ control sample. The $\BB$ corrections depend on $\egcms$; the resulting 
correlations between  the $100\mev$ \egcms bins have been taken into account in the computation of the total 
systematic uncertainty in the PBFs and moments. For example, for  $2.0 \gev < \eg < 2.7 \gev$, the $\BB$ corrections 
contribute 5.5\% to a total systematic uncertainty of 8.5\%  of the PBF, and 0.008 \gev  and 0.0009 $\gev^{2}$ of 
the total systematic uncertainty of the first and second moments, respectively. Additional contributions to the PBF  
uncertainty (added in quadrature), all energy-independent, come from the photon selection (3.3\%) due  to the 
photon efficiency, determined with $\piz$'s from $\tau$ decay, and the isolation requirement, calorimeter energy 
scale and resolution, determined from  $\B \rightarrow K^{*}\gamma$ decays and photons from virtual Compton 
scattering; efficiency of the event shape variable selection (3\%), determined  from a $\piz$ control sample; 
the semi-leptonic corrections (3\%); lepton identification (2\%) and the modeling of the $X_{s}$ fragmentation 
(1.5\%). Additional uncertainties to the first and second moment, added in quadrature,  come  from the uncertainty 
in the calorimeter energy scale (0.006 \gev) and resolution (0.0004 $\gev^{2}$), respectively.

The parameters  \mb and \mupisq ,  which are defined differently  in the kinetic (K) and shape function (SF) schemes, 
can be extracted by fitting theoretical predictions  to the measured moments. The first moments for 
$\eg > 1.9$ and $2.0 \gev$ and the  second moment for $\eg>2.0\gev$  are fitted, taking into account the correlations 
between the measured moments.  As the moments are dependent on the assumed signal model due to the efficiency and 
resolution smearing corrections,  the signal model and the model-dependence errors are adjusted based on the results 
of the fit and the moments are
recomputed and refit. Only a few iterations are required until the result is stable. In the  kinetic scheme  
$\mbkin = 4.44^{+0.08+0.12}_{-0.07-0.14}\gevcc$ and $\mupisqkin = 0.64^{+0.13+0.23}_{-0.12-0.24} \gev^{2}$, with a 
correlation of $-0.93$. The first error is due to the uncertainty in the measured moments and the second error is 
due to uncertainty in the theoretical calculations~\cite{ref:BBU}. In the shape function scheme, using the 
exponential shape function form~\cite{ref:N}, $\mbsf = 4.43^{+0.07}_{-0.08} \gevcc$ and 
$\mupisqsf =0.44^{+0.06}_{-0.07} \gev^{2}$, with a correlation of $-0.63$. If the Gaussian shape function form were 
used, $\mbsf$ and $\mupisqsf $ would increase by 0.13 \gevcc and 0.01 $\gev^{2}$, respectively. The spectra with the 
fitted parameters  are compared to data in figure~\ref{fig:spectrumuncorrected}. These results (without theory error) 
are then used to extrapolate the measured partial branching fraction from $\eg>1.9\gev$ to  1.6\gev to allow comparisons to 
theoretical predictions. In the kinetic scheme $\BR(\bxsg,\eg>1.6\gev)=(3.94\pm0.31\pm0.36\pm0.21)\times 10^{-4}$ and 
in the shape function scheme $\BR(\bxsg,\eg>1.6\gev)=(4.79\pm0.38\pm0.44^{+0.73}_{-0.47}) \times 10^{-4}$, where the 
errors are statistical, systematic and model-dependence. The model-dependence is derived from the $1\sigma$ error 
ellipse for the \mb-\mupisq fit. The central value in the shape function scheme is reduced to $4.55 \times 10^{-4}$ 
if the Gaussian form is used.

Finally the sample is divided into $b$ and $\bbar$ decays using the charge of the lepton tag to measure 
$\acp(B \to X_{s+d}\gamma) = \frac{N^{+}-N^{-}}{N^{+}+N^{-}} \frac{1}{1-2\omega}$ where $N^{+(-)}$ are the 
positively (negatively) tagged signal yields and  $1/(1-2\omega)$ is the dilution factor due to the mistag 
fraction  $\omega$. A requirement $2.2<\egcms<2.7\gev$ maximizes the statistical precision of the measurement
as determined from simulated data. The yields are  $N^{+}=349\pm48$ and $N^{-}=409\pm 45$. The bias on \acp due 
to any charge asymmetry in the detector or \BB background is measured to be $-0.005 \pm 0.013$ using control samples 
of $\epem \to X \gamma$ and $\B \to X \piz,\eta$. The mistag fraction due to mixing is $9.3 \pm 0.2\%$~\cite{ref:pdg}.
 An additional $2.6\pm 0.3\%$ mistag fraction arises from leptons from $D$ decay, $\pipm$ faking $\mu^{\pm}$, 
$\gamma$ conversions, \piz Dalitz decay, and charmonium decay. After correcting for charge bias and dilution
$  \acp  = -0.110 \pm 0.115 (\mathrm{stat.}) \pm 0.017 (\mathrm{syst.}) $, including multiplicative systematic 
uncertainties from the \BB background subtraction (5.4\%) and the dilution factor (1.0\%). The  model-dependence 
uncertainty due to differences in the \bxsg and \bxdg spectra is estimated to be negligible.

In conclusion, the branching fraction and the energy moments of the photon spectrum in  $\bxsg$ are measured 
for $\eg > 1.9 \gev$. The moments are consistent with previous measurements~\cite{ref:cleobsg,ref:bellebsg,
ref:semi-inclusive} and are used to extract values of $\mb$ and $\mupisq$ which are consistent with those 
extracted from semi-leptonic \B decays~\cite{ref:BaBarVbc}. These measurements have been used to reduce the 
systematic error in the estimation of $\Vcb$ and $\Vub$~\cite{ref:henning}. The measured branching fractions are 
in  agreement with the SM expectation and previous measurements. The measured \acp is also consistent with the 
SM expectation.

\input acknow_PRL.tex

\appendix
\begin{flushleft}
{\bf Appendix}

\input appendix

\end{flushleft}

\end{document}

%% file: bsg_symbols.tex
\input babarsym
\newcommand{\etal}      {\mbox{\textsl{et al.}}\xspace}

 \def\xs         {\ensuremath{X_{s} }\xspace}
 
 \def\pizeta     {\ensuremath{\piz(\eta)}\xspace}

 \def\eg         {\ensuremath{E_{\gamma} }\xspace}

 \def\egcms      {\ensuremath{E^{*}_{\gamma}}\xspace}

 \def\mxs        {\ensuremath{m_{X_{s}} }\xspace}

 \def\mup        {\ensuremath{\mu_{\pi}}\xspace}
 \def\mupisq        {\ensuremath{\mu_{\pi}^{2}}\xspace}
\def\mupisqkin        {\ensuremath{\mu_{\pi(K)}^{2}}\xspace}
\def\mupisqsf        {\ensuremath{\mu_{\pi(SF)}^{2}}\xspace}
 \def\mb         {\ensuremath{m_{b} }\xspace}  
 \def\mbkin         {\ensuremath{m_{b(K)} }\xspace}  
 \def\mbsf         {\ensuremath{m_{b(SF)} }\xspace}  
 
 \def\acp        {\ensuremath{A_{CP}}\xspace}

 \def\bsg        {\ensuremath{b \to s \gamma}\xspace}
 \def\bdg        {\ensuremath{b \to d \gamma}\xspace}

 \def\bxsg       {\ensuremath{B \to X_{s} \gamma}\xspace}
 \def\bxdg       {\ensuremath{B \to X_{d} \gamma}\xspace}

\def\efmbrest   {\ensuremath{\langle \eg \rangle}\xspace}

\def\esmbrest   {\ensuremath{\langle \eg^{2} \rangle}\xspace}

\def\varbrest   {\ensuremath{\esmbrest - {\efmbrest}^{2}}\xspace}

\def\aveDelta#1 {\ensuremath{\langle \Delta_{total}#1 \rangle}\xspace}

%% file: authors.tex
%
\author{B.~Aubert}
\author{R.~Barate}
\author{M.~Bona}
\author{D.~Boutigny}
\author{F.~Couderc}
\author{Y.~Karyotakis}
\author{J.~P.~Lees}
\author{V.~Poireau}
\author{V.~Tisserand}
\author{A.~Zghiche}
\affiliation{Laboratoire de Physique des Particules, F-74941 Annecy-le-Vieux, France }
% ...
\collaboration{The \babar\ Collaboration}
\noaffiliation

%% file: results.tex
\begin{table*}[hbt]
\hspace*{1.0cm}
 \begin{center}
\caption{The measured partial branching fraction, first and second moment ($\pm stat. \pm syst. \pm model$) for different ranges of $\eg$ in the B rest frame. }
  \begin{tabular}{|c|c|c|c|c|} \hline
   $\eg~(\mathrm{GeV})$      & $\Delta\BR(\bxsg)~(10^{-4})$                &   $\efmbrest~(\mathrm{GeV})$                             &  $\varbrest~(\mathrm{GeV}^{2})$        \\ \hline
   1.9 to 2.7   & $3.67 \pm 0.29 \pm 0.34 \pm 0.29$    &  $2.288 \pm 0.025 \pm 0.017 \pm 0.015$   &  $0.0328 \pm 0.0040 \pm 0.0023 \pm 0.0036$   \\       
   2.0 to 2.7   & $3.41 \pm 0.27 \pm 0.29 \pm 0.23$    &  $2.316 \pm 0.016 \pm 0.010 \pm 0.013$   &  $0.0266 \pm 0.0026 \pm 0.0010 \pm 0.0020$   \\       
   2.1 to 2.7   & $2.97 \pm 0.24 \pm 0.25 \pm 0.17$    &  $2.355 \pm 0.014 \pm 0.007 \pm 0.011$   &  $0.0191 \pm 0.0019 \pm 0.0006 \pm 0.0015$   \\       
   2.2 to 2.7   & $2.42 \pm 0.21 \pm 0.20 \pm 0.13$    &  $2.407 \pm 0.012 \pm 0.005 \pm 0.008$   &  $0.0116 \pm 0.0014 \pm 0.0004 \pm 0.0005$   \\ \hline
  \end{tabular}
 \end{center}
 \vspace{-0.1in}
 \label{tab:results}
\end{table*}

%% file: acknow_PRL.tex
We are grateful for the excellent luminosity and machine conditions
provided by our \pep2\ colleagues, 
and for the substantial dedicated effort from
the computing organizations that support \babar.
The collaborating institutions wish to thank 
SLAC for its support and kind hospitality. 
This work is supported by
DOE
and NSF (USA),
NSERC (Canada),
IHEP (China),
CEA and
CNRS-IN2P3
(France),
BMBF and DFG
(Germany),
INFN (Italy),
FOM (The Netherlands),
NFR (Norway),
MIST (Russia),
MEC (Spain), and
PPARC (United Kingdom). 
Individuals have received support from the
Marie Curie EIF (European Union) and
the A.~P.~Sloan Foundation.

%% file: appendix.tex
The correlation matrices for the statistical, systematic and model-dependence errors of the
first and second moments are given in tables~\ref{tab:moment_correlations_stat},~\ref{tab:moment_correlations_syst}
and ~\ref{tab:moment_correlations_model} respectively.  The matrices are symmetric so only the upper half is tabulated. 
The moments are measured for four energy ranges, $1.9$,$2.0$,$2.1$,$2.2$  $< \eg < 2.7 \gev$. 

\begin{table*}[htp]
\hspace*{1.0cm}
 \begin{center}

  \caption{The correlation matrix for the statistical errors between  the
   the  measured moments. FM and SM denote first and second moment respectively.}
  \vspace{0.1in}
  \begin{tabular}{|rc|cccccccc|}
   \hline
                                 &&  FM  &  FM  &  FM  &  FM  &  SM  &  SM  &  SM  &  SM  \\
   Quantity & Min. \eg  & 1.9 & 2.0 & 2.1 & 2.2 & 1.9 & 2.0 & 2.1 & 2.2 \\ \hline
   FM     & $1.9\gev$ & 1.0000 & 0.5172 & 0.3548 & 0.2265 &-0.6110 & 0.0077 & 0.0821 & 0.1008  \\ 
   FM     & $2.0\gev$ &        & 1.0000 & 0.6838 & 0.4326 & 0.2285 &-0.0008 & 0.1375 & 0.1645  \\
   FM     & $2.1\gev$ &        &        & 1.0000 & 0.6260 & 0.4220 & 0.5660 & 0.1650 & 0.1884  \\     
   FM     & $2.2\gev$ &        &        &        & 1.0000 & 0.4528 & 0.7568 & 0.7383 & 0.2113  \\     
   SM     & $1.9\gev$ &        &        &        &        & 1.0000 & 0.4626 & 0.3486 & 0.2520  \\  
   SM     & $2.0\gev$ &        &        &        &        &        & 1.0000 & 0.6966 & 0.4887  \\
   SM     & $2.1\gev$ &        &        &        &        &        &        & 1.0000 & 0.6709  \\
   SM     & $2.2\gev$ &        &        &        &        &        &        &        & 1.0000  \\
   \hline
  \end{tabular}
  
  \label{tab:moment_correlations_stat}
 \end{center}
\end{table*}

\vspace{-0.3in}

\begin{table*}[htp]
 \begin{center}
  \caption{The correlation matrix for the systematic errors between  the
   the  measured moments. FM and SM denote first and second moment respectively.}
  \vspace{0.1in}
  \begin{tabular}{|rr|cccccccc|}
   \hline
                                 &&  FM  &  FM  &  FM  &  FM  &  SM  &  SM  &  SM  &  SM  \\
   Quantity & Min. \eg  & 1.9 & 2.0 & 2.1 & 2.2 & 1.9 & 2.0 & 2.1 & 2.2 \\ \hline
   FM     & $1.9\gev$ & 1.0000 & 0.7841 & 0.6076 & 0.4622 &-0.7552 &-0.3862 & -0.1723 & -0.0942  \\
   FM     & $2.0\gev$ &        & 1.0000 & 0.8007 & 0.6311 &-0.2515 &-0.4753 & -0.2298 & -0.1451  \\   
   FM     & $2.1\gev$ &        &        & 1.0000 & 0.7886 &-0.0320 & 0.0202 & -0.2903 & -0.2115  \\
   FM     & $2.2\gev$ &        &        &        & 1.0000 & 0.0489 & 0.1850 &  0.2061 & -0.1001  \\  
   SM     & $1.9\gev$ &        &        &        &        & 1.0000 & 0.4473 &  0.2219 &  0.1436  \\
   SM     & $2.0\gev$ &        &        &        &        &        & 1.0000 &  0.5003 &  0.3342  \\
   SM     & $2.1\gev$ &        &        &        &        &        &        &  1.0000 &  0.6592  \\
   SM     & $2.2\gev$ &        &        &        &        &        &        &         &  1.0000  \\
   \hline
  \end{tabular}
  \label{tab:moment_correlations_syst}
 \end{center}
\end{table*}

\vspace{-0.3in}

\begin{table*}[htp]
 \begin{center}
  \caption{The correlation matrix for the model-dependence errors between  the
   the  measured moments. FM and SM denote first and second moment respectively.}
  \vspace{0.1in}
  \begin{tabular}{|rr|cccccccc|}
   \hline
                                 &&  FM  &  FM  &  FM  &  FM  &  SM  &  SM  &  SM  &  SM  \\
   Quantity & Min. \eg  & 1.9 & 2.0 & 2.1 & 2.2 & 1.9 & 2.0 & 2.1 & 2.2 \\ \hline
   FM     & $1.9\gev$ & 1.0000 & 0.9267 & 0.9486 & 0.8252 &-0.9057 &-0.9223  &-0.9234 &-0.7983  \\
   FM     & $2.0\gev$ &        & 1.0000 & 0.9576 & 0.8669 &-0.7452 &-0.8183  &-0.8087 &-0.8960  \\   
   FM     & $2.1\gev$ &        &        & 1.0000 & 0.9540 &-0.7390 &-0.7889  &-0.7843 &-0.7752  \\
   FM     & $2.2\gev$ &        &        &        & 1.0000 &-0.5378 &-0.5857  &-0.5788 &-0.6023  \\  
   SM     & $1.9\gev$ &        &        &        &        & 1.0000 & 0.9650  & 0.9824 & 0.6983  \\
   SM     & $2.0\gev$ &        &        &        &        &        & 1.0000  & 0.9810 & 0.8035  \\
   SM     & $2.1\gev$ &        &        &        &        &        &         & 1.0000 & 0.8057  \\
   SM     & $2.2\gev$ &        &        &        &        &        &         &        & 1.0000  \\
   \hline
  \end{tabular}
  \label{tab:moment_correlations_model}
 \end{center}
\end{table*}